\documentstyle[graphics]{aipproc}
\newcommand{\bc}{\begin{center}}
\newcommand{\ec}{\end{center}}
\newcommand{\ba}{\begin{array}}                         
\newcommand{\ea}{\end{array}}
\newcommand{\be}{\begin{equation}}
\newcommand{\ee}{\end{equation}}
\newcommand{\bea}{\begin{eqnarray}}                    
\newcommand{\eea}{\end{eqnarray}}
\newcommand{\bfig}{\begin{figure}}
\newcommand{\efig}{\end{figure}}
\newcommand{\btb}{\begin{table}}
\newcommand{\etb}{\end{table}}
\newcommand{\btbr}{\begin{tabular}}
\newcommand{\etbr}{\end{tabular}}
\newcommand{\bqo}{\begin{quote}}
\newcommand{\eqo}{\end{quote}}

\newcommand{\bfx}{{\bf x}}
\newcommand{\bfF}{{\mathbf{F}}}

\newcommand{\bfesi}{{\mathbf{e}}_{s,i}}
\newcommand{\bffsi}{{\mathbf{f}}_{s,i}}
\newcommand{\bfeui}{{\mathbf{e}}_{u,i}}
\newcommand{\bffui}{{\mathbf{f}}_{u,i}}
\newcommand{\bff}{{\mathbf{f}}}
\newcommand{\bfg}{{\mathbf{g}}}
\newcommand{\bfU}{{\mathbf{U}}}
\newcommand{\bfV}{{\mathbf{V}}}
\newcommand{\bfW}{{\mathbf{W}}}
\newcommand{\bfX}{{\mathbf{X}}}

\newcommand{\dpmax}{\delta p_{\rm max}}
\newcommand{\dpn}{\delta p_n}
\begin{document}
\thispagestyle{empty}
\newpage
\title{The Control of Dynamical Systems\\
-- Recovering Order from Chaos --}

\author{Louis J. Dub\'e and Philippe Despr\'es}
\address{D\'epartement de Physique, Universit\'e Laval\\
Cit\'e Universitaire, Qu\'ebec, Canada G1K 7P4}

\maketitle

\begin{abstract}
 Following a brief historical introduction of the notions of chaos in dynamical systems, we will present recent developments that attempt to profit from the rich structure and complexity of the chaotic dynamics. In particular, we will demonstrate the ability to 
{\bf control chaos} in realistic complex environments. Several  applications will serve to illustrate the theory and to highlight its advantages and weaknesses. 
The presentation will end with a survey of possible generalizations
and extensions of the basic formalism as well as a discussion of
applications outside the field of the physical sciences.
Future research avenues in this rapidly growing field will also 
be addressed.
 
\end{abstract}
\vfill
\begin{center}
-- Octobre 1999 --
\end{center}
\vfill
\footnotesize{
Invited Talk at the XXIth International Conference on the Physics of Electronic and Atomic
Collisions (ICPEAC), July 22-27, 1999 (Sendai, Japan).\\
{\ }\\
in {\em The Physics of Electronic and Atomic Collisions},
ed. Y. Itikawa, AIP Conference Proceedings {\bf 500}, 551-570 (2000) (AIP: Woodbury, N.Y.) }

\clearpage
\pagestyle{plain}
\setcounter{page}{1}
\title{The Control of Dynamical Systems\\
-- Recovering Order from Chaos --}

\author{Louis J. Dub\'e and Philippe Despr\'es}
\address{D\'epartement de Physique, Universit\'e Laval\\
Cit\'e Universitaire, Qu\'ebec, Canada G1K 7P4}

\maketitle

\begin{abstract}
 Following a brief historical introduction of the notions of chaos in dynamical systems, we will present recent developments that attempt to profit from the rich structure and complexity of the chaotic dynamics. In particular, we will demonstrate the ability to 
{\bf control chaos} in realistic complex environments. Several  applications will serve to illustrate the theory and to highlight its advantages and weaknesses. 
The presentation will end with a survey of possible generalizations
and extensions of the basic formalism as well as a discussion of
applications outside the field of the physical sciences.
Future research avenues in this rapidly growing field will also 
be addressed.
 
\end{abstract}

\section*{Introduction}
In his  1985 Gifford Lectures, Freeman Dyson expressed his opinion on the matter
of chaos. In his subsequently published words \cite{DYS88}, the chapter entitled
``Engineers' Dreams'' contains the following statement:
\begin{center}
{\footnotesize
\begin{quote}
 A {\bf chaotic motion} is generally neither predictable nor controllable.
It is {\bf unpredictable} because a small disturbance will produce exponentially
growing perturbation of the motion.
It is {\bf uncontrollable} because small disturbances lead only to other chaotic 
motions and not to any stable and predictable alternative.
Von Neumann's mistake was to imagine that every unstable motion could be nudged
into a stable motion by small pushes and pulls applied at the right places.
\end{quote} }
\end{center}
As one can see, the assertion was also meant as an answer to comments made 
by von Neumann in the early 1950s and who obviously held a less pessimistic point of view.
Dyson's position represents well the traditional wisdom until 1990 ...

In order to appreciate why the juxtaposition of the two words chaos and control is so 
counter-intuitive and why  Dyson's statement was so representative and sensible,
an operational definition of chaos will be helpful.   In our exposition, the word  chaos
has a  technical and precise meaning to be distinguished from its greek origin 
where it designated  ``the primeval emptiness of the universe before things came into being
of the abyss of Tartarus, the underworld" . A universal  
definition is not available, but most researchers would agree that {\em deterministic chaos}
could be described as follows: \\
{\em Chaos is a  long-term aperiodic behaviour of a  dynamical system
that possesses the property of  sensitivity to initial conditions.}\\
-- long-term aperiodic behaviour
   means that the time evolution of the system does not tend towards a
   stationary or periodic state, i.e. regularity of the motion is absent.\\ 
--  dynamical system
   describes a process whose future behaviour is strictly determined 
   by its past state, i.e. determinism is present and  the source of the irregularity is inherent
   and not to be found in a stochastic component.\\  
--  sensitivity to initial conditions
  implies that a very small deviation in the initial conditions  
  is sufficient to create large deviations in the future states (the so-called  ``butterfly effect''), i.e.
  despite the presence of determinism,  practical long-term predictability is lost.  

This is the type of motion that Dyson had in mind. It is not new of course and it is
clear that Maxwell and Boltzmann, the founders of statistical physics,  were acutely aware of the property of sensitivity to initial
conditions and its consequences. Not before Poincar\'e \cite{POI1890} however, could one ascertain
the existence of this property in a system with {\em few} degrees of freedom, namely the reduced 3-body problem.  
In the continuing history of nonlinear dynamical systems, the first evidence of physical chaos is 
associated with the name of Edward Lorenz \cite{LOR63}, whose 1963 discovery of the first {\em strange
attractor}  in a simplified meteorological model containing only 3 state variables has led to a remarkable
explosion in the study of chaos and its properties.  It was not until 1990 however that 
Ott, Grebogi and Yorke (OGY) \cite{OTT90} addressed the question of control of chaos and described,
very much in the  spirit of von Neumann, the theoretical steps necessary to achieve this goal.  This work
was rapidly followed by experimental verification \cite{DIT90}: von Neumann's
dream had become reality.

This Progress Report will describe some practical implementations for the recovery of  order from chaos.
The theoretical  foundations of the methods will be first explained for  1D and 2D systems and 
then demonstrated with some recent calculations taken from mathematics and physics.
Our conclusions and a glimpse at future applications and extensions make the last part of the
presentation.  Lack of space precludes completeness,  and the interested
reader may wish to consult the special 1997 Decembre issue of  {\it Chaos}  
for further information\footnote{As an indication of the rapid  growth
of the literature,   approximately 1500 articles have been published on the subject of Control and
Synchronization between 1990-1999.}.    

\section*{The Control Strategy}
\begin{flushright}
{\footnotesize {\em All stable processes, we shall predict.\\
All unstable processes, we shall control.}\\
{\sc John von Neumann}, circa 1950}
\end{flushright} 

In this Section, we  show how the richness, the complexity and the sensitivity
of chaotic dynamics can be used to select and stabilize at will, with small programmed perturbations,
an otherwise unstable state of the natural dynamics. The goal is to achieve this feat {\em without}
altering appreciably the original system.  It is precisely the properties that differentiate a chaotic motion from 
an irregular or unstable behaviour that are the solution to the control task. The important ingredients are:\\
-- unstable periodic orbits (UPO) are typically dense in the chaotic 
   attractor of dissipative systems or in the  stochastic web of  conservative systems, i.e. there are
   practically an infinity of unstable states to choose from. \\
-- chaotic motion is ergodic, meaning that a chaotic trajectory will revisit infinitely often
    the neighbourhood of any point within the available phase space.\\
-- chaotic dynamics is sensitive to initial conditions, implying that {\em small} perturbations will 
    naturally induce large effects.     

To go beyond qualitative description, we establish some working conditions: 
\begin{enumerate}
\item   we suppose that the dynamics can be represented by a $d$-dimensional nonlinear map
          (either given explicitly or reconstructed from the observations)
\be
     \bfx_{n+1} = \bfF(\bfx_{n}, p)
\ee
where $ p$ is an accessible system parameter, the {\em control parameter}.

\item  there exists one or more  specific UPOs for a given {\em nominal} value $p_0$
          of the parameter, defined by
\be
    \{\bfx(i,p_0) :  \bfx(i,p_0)  = \bfF^{(m)}(\bfx(i,p_0),p_0) \ , \qquad \forall \  i = 1,m \}
\ee
for an orbit of period $m$, 
    around which one wishes to stabilize the dynamics.

\item control is activated only when points of the trajectory $\{ \bfx_{n} \}$ fall in
         a small  neighbourhood of the UPOs,  usually taken to be a ball ${\cal B}_{\delta}$ 
         of radius $\delta$ around $\{ \bfx(i,p_0) \}$ ,
\be
         |\bfx_n - \bfx(i,p_0)| \le \delta  \qquad \qquad \mbox{for some} \quad i = 1,m \quad ,
\ee
 hereafter referred to as  the {\em control} or {\em $\delta-$neighbourhood}.
 
\item  we restrict the parameter variations $\delta p$, necessary to achieve
         control, to  a maximum small perturbation 
\be
      | \delta p| \le  |\delta p_{\rm max}|  \ll |p|
\ee
defining the {\em control range}.
     
\item since the position  of a periodic orbit is a function of $p$,  and we assume that the local dynamics 
         does not vary much within $|\delta p|$, a linear representation of the dynamics 
         is possible.     
\end{enumerate}
Obviously, the control range and neighbourhood are not independent and experience shows
that a judicious choice is to take them of the same order of magnitude.  
%
\subsection*{1D Control}

In a 1D system where the nonlinear map is given explicitly by
\be
       x_{n+1} = F(x_n,p) \quad ,
\label{eq-1d}
\ee
and where a target UPO of period $m$, i.e. $\{ x(i,p_0) \}$, exists at nominal value $p_0$
of the control parameter, the perturbations necessary to stabilize the orbit can be calculated 
directly. Indeed, assume that at the $n$-th iteration, $x_n$ comes in ${\cal B}_{\delta}$ of the $i$-th component of the target UPO, i.e. $|x_n - x(i,p_0)| \le \delta$, then  
equation (\ref{eq-1d})  can easily be linearized around $x(i,p_0)$ and $p=p_0$ such that
\be
        x_{n+1} - x(i+1,p_0) \sim   U_i\  [ x_n -x(i,p_0) ] + V_i\  [p_n -p_0] 
                                      \equiv U_i\ \delta x_{n,i} + V_i\ \delta p_n
\label{eq-linear1d}
\ee
where $U$ is the Jacobian of the map and $V$ expresses the parametric variation of the map
\be
      U = D_x F(x,p) = \frac{\partial}{\partial x} F(x,p)   \qquad
      V = D_p F(x,p) = \frac{\partial}{\partial p} F(x,p)   \quad .
\ee
The notation $U_i$ and $V_i$ indicates that the partial derivatives are evaluated at
$[x= x(i,p_0), p=p_0]$. To obtain an expression for $\delta p_n$, one imposes
the {\em control criterion} (not unique) that equation (\ref{eq-linear1d}), taken as a strict equality, should be
equal to zero, namely that
\be
     x_{n+1} - x(i+1,p_0) = 0  \quad .
\ee 
\hspace*{\fill} {\bf (control criterion 1D)}   \quad \quad \\
Solving for $\delta p_n$, equation (\ref{eq-linear1d}) leads immediately to 
\be
       \delta p_n = - \frac{U_i}{V_i}\  \delta x_{n,i} \quad .
 \ee
In order to complete the procedure, one should make sure that  $|\dpn|$ just
obtained is $\le \dpmax$ (we will always take $\dpmax > 0$ ).  If it is so,
$p_n = p_0  + \dpn$ for the next iteration; if not, one could set $p_n= p_0$
and wait until the trajectory reenters ${\cal B}_{\delta}$.  We have found that a more
robust choice is to apply the corrections
during the control stage according to the prescription
\be
       \dpn  \longrightarrow  \dpmax  \tanh (\dpn / \dpmax ) \qquad {\rm for} \qquad
        |\dpn| > \dpmax \quad .
\ee
This is a minor point however since typically  $\dpn$ decreases rapidly after the first
few iterations as we are getting ever closer to the target orbit ($\dpn \propto \delta x_n$).

%
\subsection*{2D Control}
 
Stabilization in higher dimensions is qualitatively different from the 1D case  
since  phase space is endowed with a much richer structure. For simplicity, we will confine
our discussion to two dimensions.  In 2D, the generic local neighbourhood of a UPO is equipped
with  a stable and unstable manifold. A chaotic trajectory entering the neighbourhood will move 
toward the UPO along the stable direction and escape along the unstable one.
This is the ``saddle dynamics'' illustrated in Figure (1).  
%
%
\begin{figure}[b]
\hfill \scalebox{0.7}{\includegraphics{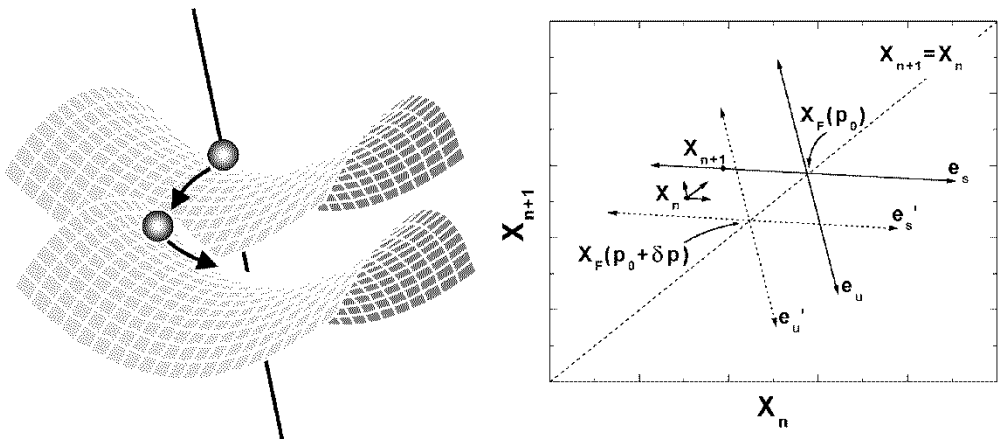}} \hfill
\vspace{10pt}
\caption{{\bf Local Geometry of Control}: ({\em left})  2D ``saddle dynamics'' and 
({\em right}) linearization of the stable and unstable manifolds.}
\end{figure}

OGY \cite{OTT90} realized that a possible solution of controlling  chaos could be obtained by locally
displacing the manifolds such that the component of the motion along the unstable direction 
could be eliminated (at least to first order) such that subsequent evolution will naturally lead the
orbit to the unstable point along the stable direction.  This idea is geometrically presented in
Figure (1)  for the situation of an unstable fixed point denoted $\bfx_F(p_0)$, much like
the task of bringing a ball bearing to rest on a saddle.  The remaining part 
of this subsection is devoted to the mathematical translation  of the applications of ``small pushes and pulls
applied at the right places''.

Starting from the 2D nonlinear map
\be
     \bfx_{n+1} = \bfF(\bfx_{n}, p) 
\ee
with  $\bfx_{n} \in {\cal R}^2$ and  $p \in {\cal R}$, one traces the steps
to achieve control through the following algorithm.
\begin{enumerate}
\item  locate a UPO of period $m$ for the nominal value of the parameter $p_0$
\be
 \bfx(1,p_0) \to \bfx(2,p_0) \to \cdots
  \bfx(m,p_0) 
  \to \bfx(m+1,p_0) = \bfx(1,p_0) \quad .
\ee

\item  linearize the dynamics  in the $\delta$-neighbourhood of
          $\bfx(i,p_0)$ :
\be
   \bfx_{n+1} - \bfx(i+1,p_0) \sim 
{\bf U}_i\ \left[ \bfx_{n} - \bfx(i,p_0) \right] + \bfV_i \ \delta p_n
\label{eq-linear2d}
\ee
where ${\bf U}$ is the $ 2 \times 2$ Jacobian matrix and $\bfV$ is a $2 \times 1$ parametric
variation vector 
\be
   {\bf U} \equiv D_{\bfx} \ \bfF(\bfx,p)   \qquad {\rm and} \qquad  
   \bfV \equiv D_p \ \bfF(\bfx,p)   
\ee
with the partial derivatives evaluated at $[\bfx = \bfx(i,p_0) , \  p = p_0]$.

\item  characterize the local dynamics by the {\em stable} $\bfesi$
          and the {\em unstable} $\bfeui$ directions.
%
%

\item  construct the contravariant vectors defined by
\be
        \bffui \cdot \bfeui = \bffsi \cdot \bfesi = 1 \qquad \qquad 
        \bffui \cdot \bfesi = \bffsi \cdot \bfeui = 0 \quad .
\ee

\item stabilize the orbit by demanding that it falls, on the next
          iteration, on the stable direction, i.e.
\be
   \bff_{u,i+1} \cdot \left[ \bfx_{n+1} - \bfx(i+1,p_0) \right]  = 0 \quad .
   \label{eq-project1}    
\ee
\hspace*{\fill} {\bf (control criterion OGY)} \quad \quad \\ 
Therefore, according to step 2, we obtain the relation
\be
 \bff_{u,i+1} \cdot 
\left\{ {\bf U}_i\ \left[ \bfx_{n} - \bfx(i,p_0) \right] + \bfV_i \ \delta p_n \right\}  = 0 \quad . 
\label{eq-project2}
\ee

\item calculate the perturbation necessary to satisfy equation (\ref{eq-project2})
\be
    \delta p_n = -\  \frac{  \bff_{u,i+1}  \cdot \left\{ {\bf U}_i\ \left[ \bfx_{n} - \bfx(i,p_0) \right]
                                \right\}  }
                      {  \bff_{u,i+1} \cdot \bfV_i   }
\ee

and apply only if $| \delta p_n| < \delta p_{\rm max} $ otherwise set e.g. $ \delta p_n= 0$ or use
equation (10).
 
\end{enumerate} 

In summary, the stabilization procedure can be divided in three separate stages:
the {\em learning stage}, where one identifies the desired UPOs, extracts the
Jacobian matrices,  and calculates the
corresponding stable and unstable directions $\bfesi$ , $\bfeui$ to construct 
the contravariant vectors   $\bffui$ ; 
the {\em transient stage}, where, after randomly choosing an initial condition,
the system is let to evolve freely at the nominal parameter value   $p_0$
until, at the {\em control stage}, once the chaotic trajectory has entered the 
prescribed $\delta$-neighbourhood, the control  is attempted  by means of small
parameter perturbations. 

%
\subsection*{Alternative Control Algorithms}

The algorithm just described is {\em dynamically optimal} in that it uses a control criterion
and perturbations  that involve the complete local structure of the system's dynamics. 
From a practical (experimental) point of view, it is also most demanding considering that the
underlying dynamical law is often not even known {\em a priori}. It requires the determination
of the Jacobian matrices and  the parametric variation of the map along the UPOs and the
corresponding stable and unstable directions.  Several modifications
of the original method have been proposed and we now present some of them with an emphasis
towards algorithms that are simpler and in some instances more practicable.

If instead of equation (\ref{eq-linear2d}), one writes the linearization as
\be
      \bfx_{n+1} - \bfx(i+1,p_n) \sim 
{\bf U}_i\ \left[ \bfx_{n} - \bfx(i,p_n) \right]
\label{eq-linearogy}
\ee
where the dependence of $\bfU_i$ on $p$ has been ignored, and one introduces
the parametric variation of the periodic  points as  
\be
      {\bf g}_i \equiv  \left. \frac{d}{d p}\ \bfx(i,p) \right|_{p=p_0} 
                     \sim  \frac{\bfx(i,p_0 + \delta p) - \bfx(i,p_0)}{\delta p}
\ee
or equivalently,
\be 
      \bfx(i,p_0 + \delta p) \sim  \bfx(i,p_0) +  \bfg_i\ \delta p \quad ,
\ee
one arrives, under the criterion (\ref{eq-project1}), to the perturbations
\be
      \delta p_n = -\  \frac{  \bff_{u,i+1}  \cdot \left\{ {\bf U}_i\ \left[ \bfx_{n} - \bfx(i,p_0) \right]
                                \right\}  }
                      {  \bff_{u,i+1} \cdot (\bfg_{i+1} - \bfU_i\ \bfg_i)   } \quad .
\label{eq-dpng}
\ee
The expression (\ref{eq-dpng}) has the advantage that the variables $\bfg_i$ can easily be obtained 
from  observations of the shift of the periodic points under small parameter change.

One further simplification arises if it is sufficient to intervene on the dynamics only once per period.
The modifications to the formula are straightforward.  Equation (\ref{eq-linearogy}) becomes
 \be
      \bfx_{n+m} - \bfx_{F,i}(p_n) \sim 
{\bf U}_i^{(m)}\ \left[ \bfx_{n} - \bfx_{F,i}(p_n) \right]
\ee
where the notation is chosen to emphasize that $ \bfx_{F,i}(p)= \bfx(i,p)$ is a fixed point
of $\bfF^{(m)}$ (the $m$ times application of the map $\bfF$). Furthermore, the Jacobian matrix
${\bf U}_i^{(m)} = D_{\bfx} \bfF^{(m)}(\bfx,p)$, evaluated at $[\bfx_{F,i}(p_0), p_0]$,
can be expanded in terms of its eigenvectors (the stable and unstable manifolds) and eigenvalues
as $\lambda_{u,i}\ \bfeui \bffui + \lambda_{s,i}\ \bfesi \bffsi$ to modify (\ref{eq-dpng}) to
\be
  \delta p_n = -\  \frac{\lambda_{u,i}}{(1 - \lambda_{u,i})}
                     \frac{  \bff_{u,i}  \cdot  \left[ \bfx_{n} - \bfx_{F,i}(p_0) \right]   }
                      {  \bff_{u,i} \cdot \bfg_{i}   } \quad .
\label{eq-dpnperi}
\ee

Until now the control criterion has not been modified, but in situations where the stable  and unstable 
manifolds may be difficult to obtain (e.g. in high dimensions) or for the sake of simplicity, one might choose
to minimize the deviation from the target orbit
instead of projecting onto the stable manifold,  i.e.
we demand that 
\be 
      || \bfx_{n+m} - \bfx_{F,i} || = {\rm minimum}  \quad ,
\label{eq-critmed}
\ee
\hspace*{\fill} {\bf (control criterion MED)} \quad \quad \\ 
where an estimate of $\bfx_{n+m}$ is given  by equations (23) and (21), namely
\be
    \bfx_{n+m} \sim  \bfx_{F,i}(p_0) 
     + {\bf U}_i^{(m)}\ \left[ \bfx_{n} - \bfx_{F,i}(p_0) \right]
     + (\bfg_i - \bfU_i^{(m)}\ \bfg_i)\ \delta p_n \quad .
\ee
The solution of the minimization (\ref{eq-critmed}) is then
\be
       \delta p_n = -\  \frac{ (\bfg_i - \bfU_i^{(m)}\ \bfg_i)   \cdot \left\{ {\bf U}_i^{(m)}\ 
                       \left[ \bfx_{n} - \bfx_{F,i}(p_0) \right] \right\}  }
                      {  ||(\bfg_i - \bfU_i^{(m)}\ \bfg_i)||^2   } \quad .
\label{eq-dpnmed}
\ee
 This technique was first introduced in \cite{REY93} and goes by the name of minimal expected deviation
(MED) method\footnote{For conciseness, we only give the FIRST reference for each technique, although
the methods are continuously being improved and modified: this applies to the entire report.}.   

The perturbations $\delta p_n$ transform the original autonomous systems to nonautonomous ones.
One could therefore consider a formulation extending phase space by one dimension with $p_n$
as the new dynamical variable.  Alternatively, as was first realized in \cite{DRE92},  one could account 
explicitly for the $p_n$ dependence by introducing in the mapping itself the past history of the
perturbations, namely 
\be
        \bfx_{n+1} = \bfF_c(\bfx_n, p_n, p_{n-1}) \quad .
\ee
We restrict ourselves to the last two perturbations. The sub-index on $\bfF_c$ is to remind us
that the mapping differs from the original one and is only identical to it when $p_n=p_{n-1}=p_0$. 
In analogy  to equation (26), one can write
\bea
    \bfx_{n+m} &\sim&  \bfx_{F,i}(p_0,p_0) 
     + {\bf U}_i^{(m)}\ \left[ \bfx_{n} - \bfx_{F,i}(p_0,p_0) \right] \nonumber \\
                      & &     + (\bfg_{u,i} - \bfU_i^{(m)}\ \bfg_{u,i})\ \delta p_n 
      + (\bfg_{b,i} - \bfU_i^{(m)}\ \bfg_{b,i})\ \delta p_{n-1} \quad .
\label{eq-linearrpf}
\eea
with 
\be
     \bfg_u = \left. \frac{d}{d p}\ \bfx_F(p,p') \right|_{p=p'=p_0} \qquad \qquad
     \bfg_b = \left. \frac{d}{d p'}\ \bfx_F(p,p') \right|_{p=p'=p_0} \quad .
\ee
To complete the modification, a control condition must be imposed and a reasonable choice
is 
\be
            [\bfx_{n+2m} - \bfx_{F,i}(p_0,p_0)] = 0 \qquad {\rm and} \qquad
            \delta p_{n+1}  = 0 \quad .
\label{eq-critrpf}
\ee
\hspace*{\fill} {\bf (control criterion RPF)} \qquad \quad \\  
The conditions (\ref{eq-critrpf}) are sufficient to solve the linearized equation (\ref{eq-linearrpf}) 
for $\delta p_n$ as
\be
       \delta p_n = -\  \frac{ \bfX_i^{(m)} \cdot
                                        \{ \bfU_i^{(m)}\ \bfU_i^{(m)}\ [\bfx_n - \bfx_{F,i}(p_0,p_0)] \} }
        { ||\bfX_i^{(m)}||^2 }
                -\  \frac {\bfX_i^{(m)} \cdot (\bfU_i^{(m)}\ \bfW_i^{(m)}) }
 { ||\bfX_i^{(m)}||^2 }\
          \delta p_{n-1}
\ee
with $\bfX_i^{(m)} = \bfU_i^{(m)}\ \bfV_i^{(m)} + \bfW_i^{(m)}$,
$\bfV_i^{(m)} = \bfg_{u,i} - \bfU_i^{(m)}\ \bfg_{u,i}$ and 
$\bfW_i^{(m)} = \bfg_{b,i} - \bfU_i^{(m)}\ \bfg_{b,i}$.
Our derivation is somewhat different from \cite{ROL93} and , in view of the numerous parameters
to determine, it should be taken as a serious alternative only for 1D (or quasi 1D) systems. In the 
latter case, the method has a simple geometrical interpretation as illustrated in the right panel of
Figure (2) for an unstable fixed point $x_F(p_0)$: $U$ is the local slope
of the original map, $g_u \delta p$ and $g_b \delta p$ correspond
to the shifts of the fixed point  to positions $x_{F,u}(p_0+\delta p, p_0) \sim x_F(p_0) + g_u \delta p$
and $x_{F,b}(p_0, p_0 +\delta p) \sim x_F(p_0) + g_b \delta p$ respectively,
whereas $V \delta p$  denotes the map displacement from
$F_c(x, p_0, p_0) \equiv F(x,p_0) \to F_u(x,p_0+\delta p, p_0)$ and  $W \delta p$
from $F(x,p_0) \to F_b(x,p_0, p_0+\delta p)$.  These parameters are readily obtained 
from the observations as was first demonstrated by \cite{ROL93} who gave the algorithm the name
of recursive proportional feedback (RPF).
%
%
\begin{figure}[t]
\hfill \scalebox{0.9}{\includegraphics{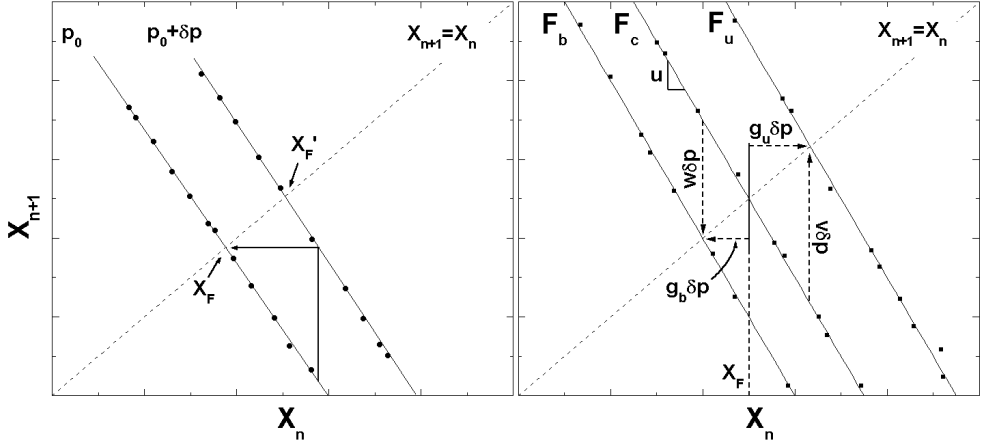}} \hfill
\vspace{10pt}
\caption{{\bf Alternative Local Geometries of Control}:
({\em left}) the Occasional Proportional Feedback (OPF) method and 
({\em right}) the  Recursive Proportional Feedback (RPF) method.}
\end{figure}

Our last modification, which we only quote for 1D systems where it is most likely to be used,
consists of ignoring in the RPF formula the dependence on $p_{n-1}$ which amounts to
setting $W=0$ in the previous equations, to obtain
\be
      \delta p_n = -\  \frac{U}{(g - U\ g)}\ [x_n - x_{F,i}(p_0)] \quad .
\label{eq-dpnopf}
\ee
One should not be deceived by the apparent simplicity of equation (\ref{eq-dpnopf}).
In its regime of applicability, it can be enormously successful as we will see shortly and
as Hunt \cite{HUN91} first demonstrated.  Its geometrical interpretation is shown in the 
left panel of Figure (2) where it amounts to choose a perturbation such that upon the next
iteration, $x_n$ is mapped onto the target fixed point. 
This occasional proportional feedback (OPF),
as it was called originally, is eminently suited for experimental control since it requires
feedback signals proportional to the deviations to the target orbit and two parameters
$U$ and $g$ readily available. Even more, if one is bold enough, one might just as well adjust
the proportionality constant until control is established.  

 \section*{Mathematical and Physical Applications}
We have selected five examples of increasing complexity and novelty.
The first three correspond to dissipative systems confined to a chaotic
attractor, whereas the last two are conservative Hamiltonian systems
where the attractor is replaced by bounded regions of phase space where the
 dynamics is chaotic.  

We will apply the perturbations only once per period for all cases except the last. 
Save for the two discrete maps, ALL the relevant control informations are obtained
numerically in an effort to simulate more closely an experimental setting.   
Reliable methods (see e.g. \cite{OTT94}) exist to locate the positions of the UPOs and we will assume 
thereafter that the locations are known prior to the control session. The numerical construction
of the Jacobian matrices from time series is often a subtle task and is beyond the scope of this article.
The reader is referred to \cite{jacobian} for technical details. 
{
  
\subsection*{1D Logistic Map}

Our first example is the {\em logistic map} 
\be
    x_{n+1} = {r}\,  x_n (1 - x_n)
\label{eq-logis}
\ee  
which is known to have a broad range of parameter values $r \in [\sim 3.57, 4.0]$ where 
chaotic motion can be observed. The variables entering the control signals are simply 
\be
        U = r (1 - 2 x)  \qquad \qquad {\rm and} \qquad \qquad V  = x ( 1 - x)
\ee
and for a given UPO, $\{ x(i) \}$, the perturbations are
\be
        \dpn = r_0 \frac{2x(i) -1}{x(i) (1 -x(i) )} \ [x_n - x(i)]
\label{eq-dpnlog}
\ee
for $|x_n - x(i)| \le \delta \ll 1$. 

%
\begin{figure}[h]
\hfill \scalebox{0.7}{\includegraphics{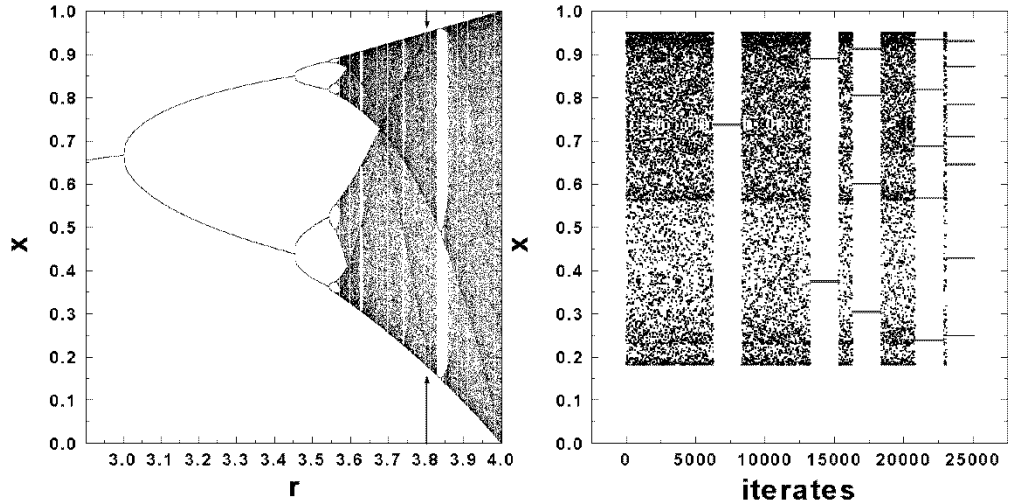}} \hfill

\vspace{10pt}
\caption{{\bf Logistic Map}: The left panel shows the bifurcation diagram as a function of
the parameter $r$. The right panel illustrates the successive control of UPOs 
of period 1, 2, 4, 5, 7
for $r_0= 3.8$ (indicated by the arrow in the left panel). 
The control is held for 2 000 iterates with $\delta = 10^{-4}$ and $\delta r_{\rm max} = 10^{-3}\ r_0$. }
\end{figure}

The left panel of Figure (3) shows the asymptotic behaviour
of the orbits for different values of $r$.  We have chosen the nominal value, $r_0 = 3.8$,
and control settings of
$\delta = 10^{-4}$, $\delta r_{\rm max} = 10^{-3}\ r_0$, and requested from our controller 
to successively stabilize periods $m= 1, 2, 4, 5, 7$ and hold control for 2 000 iterates each.
The right panel of Figure (3) shows the  flexibility of the procedure letting ergodicity
bring the trajectory near the next target orbits after a successful control sequence.  

A few remarks are in order. First, note
that the control process does not create the UPOs, they exist already in the natural (free) dynamics
and build, so to speak, the lattice upon which the chaotic trajectory wanders. The control mechanism
simply picks them out of the background. 
Second,  the transients in between controlled 
periods are of varying lengths and could be considerably reduced by optimizing the
control variables and/or by steering the chaotic orbit to the UPOs by a technique called
{\em targeting} \cite{targeting} .   

%
%
%

%
\subsection*{2D Dissipative H\'enon Map}
%
%
\begin{figure}[h]
\hfill \scalebox{0.75}{\includegraphics{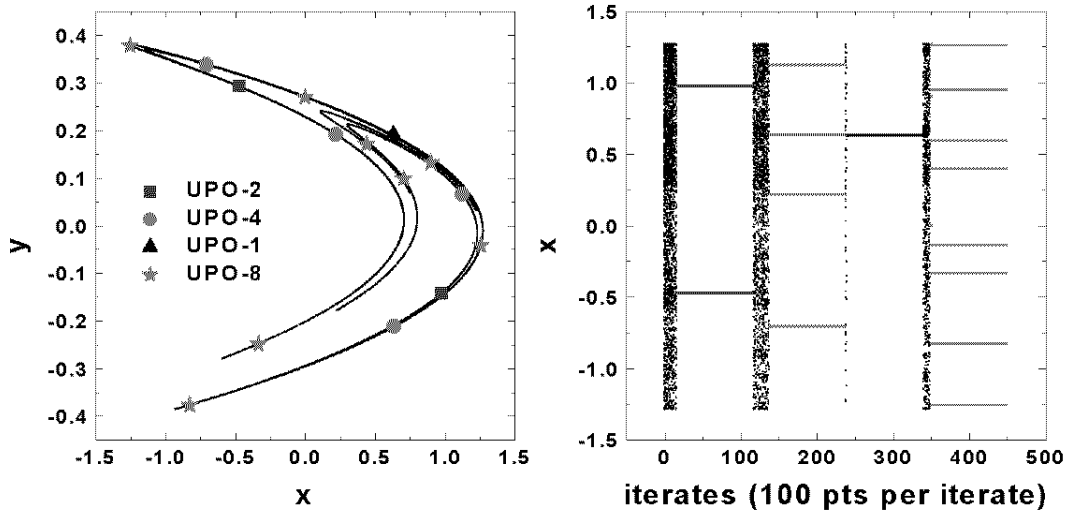}} \hfill
\vspace{10pt}
\caption{{\bf H\'enon Map}: The left panel shows the attractor with a number of embedded UPOs.
The right panel shows the successive OGY control of UPOs 
of period  2, 4, 1, 8 for $a_0 = 1.4 ,$  $b_0= 0.3$.  
The control is held for 10 000 iterates with $\delta = 5 \times 10^{-3}$ and 
$\delta a_{\rm max} = 10^{-2}\ a_0$. }
\end{figure}
The H\'enon map \cite{HEN76} has been a paradigmic example in nonlinear dynamics ever since its 
inception. It has the form  \\
\parbox{13cm}{\begin{eqnarray*} 
     x_{n+1} &=& 1 - {a}\, x_n^2 + y_n \\
     y_{n+1} &=& b\, x_n
     \end{eqnarray*} } \hfill
\parbox{1cm}{\bea \label{eq-dhenon} \eea}\\
and a Jacobian matrix given by
\be
       \bfU = \left( \begin{array}{cc}
                             -2 a  x  &  1\\
                                 b     &  0  \end{array} \right)
\ee
with determinant (the Jacobian) equal to $b$. It is dissipative for $|b| < 1$
and possesses a non trivial strange attractor for various combinations of the parameters
$a, |b| < 1$.  For our purpose, it serves as a benchmark, since the complete
OGY strategy (eqns (18) or (22) or (24) ) can be performed (semi-) analytically
and compared with other implementations.
For example, the Jacobian matrices $\bfU_i^{(m)}$ are
the product of  $m$ individual Jacobian matrices along the periodic orbit, i.e.
$\bfU_{i+m-1}\ \bfU_{i+m-2}\ \ldots\ \bfU_{i+1}\ \bfU_i$, whose eigenvectors can be written down analytically.

Our experiment (Fig. 4) shows the attractor for $a_0= 1.4, b_0= 0.3$ with the locations of
embedded UPOs of periods 1, 2, 4, 8. The OGY control algorithm was used and the perturbations
applied to the parameter $a_0$. The right panel of Figure (4) shows the values of the
$x$ variable for a scenario involving the successive stabilizations of period 2, 4, 1, and 8.  The
target orbit is changed once a UPO has been controlled for 10 000 iterates. The control region
was set to $\delta = 5 \times 10^{-3}$ and the maximum perturbation allowed was 
$\delta a_{\rm max}   = 10^{-2}\ a_0$.  Under the same operational conditions, the MED minimization
gave  equally satisfactory results.
\newpage
\subsection*{3D R\"ossler Flow}
%
%
\begin{figure}[h]
\hfill \scalebox{0.50}{\includegraphics{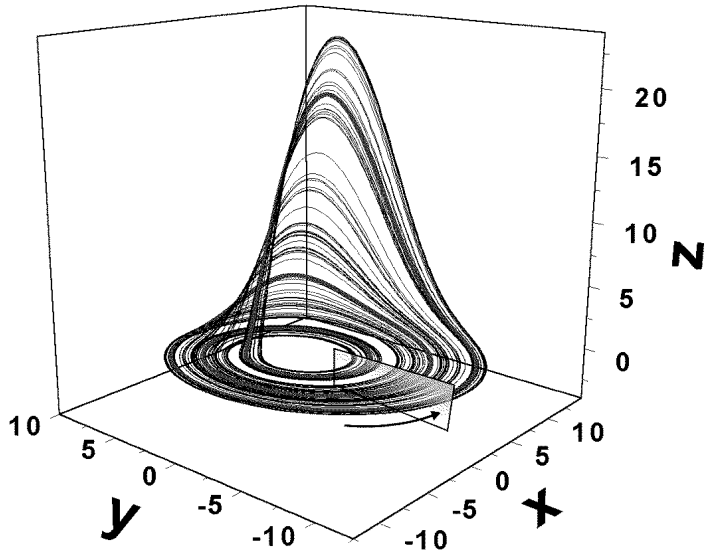}} \hfill
\vspace{10pt}
\caption{{\bf R\"ossler Attractor} with an embedded  UPO of period 3 in darker line.
The flow parameters are $a_0=b_0=0.2$ and $c_0=5.7$ .  Also indicated is the plane 
$x=2$ where dynamical informations are gathered. }
\end{figure}
Until now, our examples have described discrete dynamics and our formalism is also
derived for maps. We now show how to appropriately discretize a flow to achieve
control with the  methods derived thus far. The R\"ossler system \cite{ROS79}
consists of 3 coupled differential equations
\bea
            \dot{x} &=& -y -z \nonumber\\
            \dot{y} &=& x + a\, y  \label{eq-rossler}  \\
            \dot{z} &=& x\, z - {c}\, z +b  \nonumber
\eea
and a chaotic orbit is bounded to a funnel-like attractor (Fig. 5: $a_0=b_0=0.2, c_0=5.7$) 
where the motion is mostly in 
the $x-y$ plane with rapid excursions in the $z$ direction.

The last observation leads  us to one possible discretization: one can register
intersections of the flow with a plane $x= {\rm cte}$ (the Poincar\'e section) and accumulate
the pairs $(y_n,z_n)$ from which one could subsequently infer a map (the Poincar\'e map),
$(y_{n+1}, z_{n+1}) = \bfF_P((y_n,z_n))$. For uniqueness, one must choose {\em directed}
intersections by monitoring  $\dot{x}$  on the Poincar\'e section (see Fig. 5).

The intersections with the Poincar\'e section  ($x=2, \dot{x}>0$) is represented in the lower left panel of 
Figure (6). One notices  little variation in the $z$ component and our quasi 1D control methods
should be appropriate. In the same panel, one has indicated the positions of 3 UPOs to be stabilized.
The OPF control of the $y$ component is shown in the lower right panel of Figure (6) resulting
in 3 continuous trajectories (upper panel) embedded in the attractor (compare with Fig. 5).
We believe that this success indicates just how robust this type of linear feedback is. Remember
that we intervene in the dynamics {\em only} on the Poincar\'e section with a small perturbation
(here $< 1 \%$ of $c_0$) at every $m$ intersections.    

\newpage
%
%
\begin{figure}[t]
\hfill \scalebox{0.75}{\includegraphics{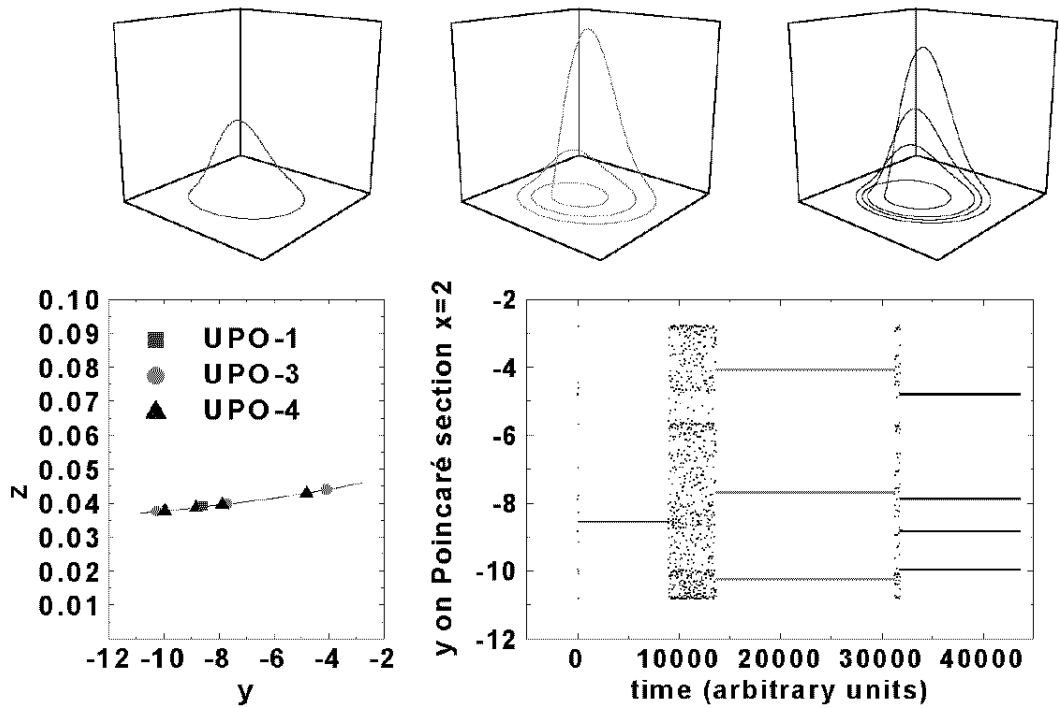}} \hfill

\caption{{\bf R\"ossler Flow}: ({\em upper panel}) 3 stabilized continuous trajectories embedded
in the attractor of Fig. 5;
({\em lower left}) locations  of the UPOs on the Poincar\'e section $x=2, \dot{x}>0$;
({\em lower right})  OPF control  on the Poincar\'e section of the $y$ components of the UPOs.
Control is held during  1 500, 1 000, and 500 cycles for period 1, 3, and 4 respectively (1 cycle = 1 complete
orbit) with $\delta = 10^{-2}$ and $\delta c_{\rm max} = 10^{-2}\ c_0$. }
\end{figure}
%
%
%
\begin{figure}[h]
\hfill \scalebox{0.75}{\includegraphics{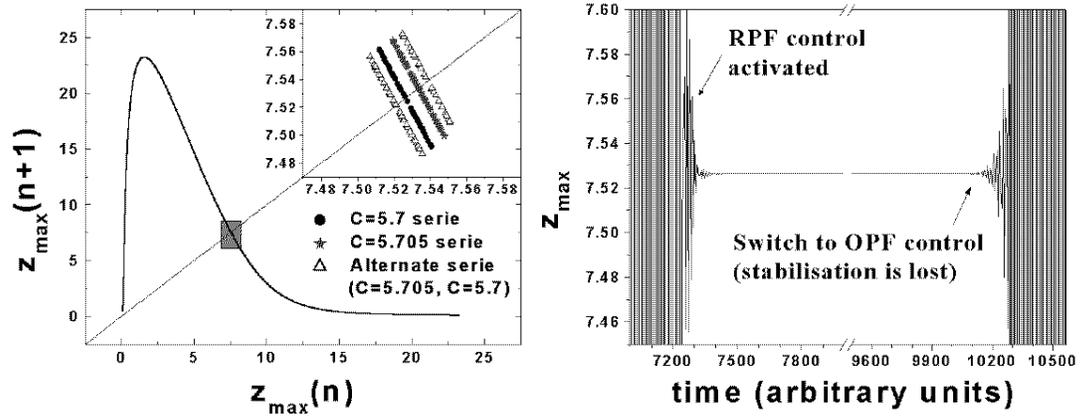}} \hfill
\caption{{\bf R\"ossler Flow}: ({\em left}) first return map of successive maxima of the
$z$ variable (Lorenz map)  for the attractor of Fig. 5;
({\em right}) success of the RPF control and failure of the OPF method applied to the $z_{\rm max}$
series with $\delta = 10^{-1}$ and $\delta c_{\rm max} = 10^{-2}\ c_0$. } 
\end{figure}
\clearpage
We have performed another experiment on the R\"ossler flow. It consists of collecting
subsequent maxima $z_{n,{\rm max}}$ of the $z$ variable along a chaotic trajectory to construct
a discretization of the flow.  By plotting $z_{n+1,{\rm max}}$ versus $z_{n,{\rm max}}$, one obtains
a  (first) return map which is usually called a Lorenz map after the man who first defined the procedure.  
Our map is shown on the left of Figure (7) and it has all the characteristics of a chaotic 1D map.
Again this indicates that our quasi 1D control (RPF or OPF) methods should be applicable. However,
according to Figure (5), this should be rather delicate since the trajectory spends most of its time in the
$x-y$ plane.  In other words, the Lorenz map may not be adequate to gather enough 
dynamical information for control. 

We have therefore used  the RPF algorithm where $g_u$ and $g_b$
are first estimated by alternating the value of $c_0$ between 5.700 and 5.705. As expected, this
creates two new applications $F_u$ and $F_b$ as clearly seen in the enlarged section of Figure (7).
The right side of Figure (7) shows the RPF stabilization of a period 1 UPO which is maintained for
5 000 iterates before switching to the OPF algorithm where control is rapidly lost. The refinement
incorporated in the RPF method has served us well and this experiment reveals nicely the
limitation  of the simplest strategy.
\subsection*{Billiard Dynamics}
%
%
\begin{figure}[b]
\hfill \scalebox{0.75}{\includegraphics{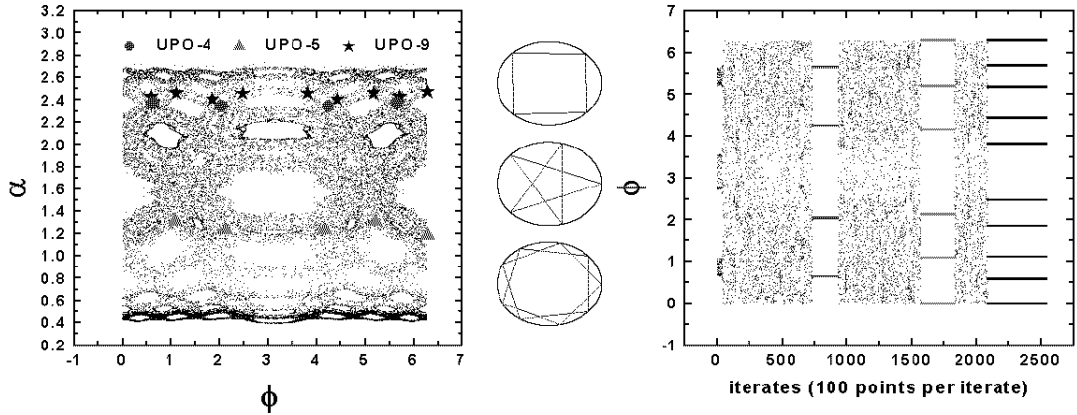}} \hfill
\vspace{10pt}
\caption{{\bf Cosine Billiard}: ({\em left}) mixed chaotic (filled) and regular 
(open islands) phase space with embedded UPOs for $\epsilon_0 = 0.3$;
({\em middle}) MED controlled UPOs of period 4, 5 , 9;
({\em right}) stabilized $\phi$ variable of the corresponding UPOs held for 5 000 cycles
each with $\delta = 10^{-2}$ (1 cycle = 1 complete orbit).}
\end{figure}

The study of the frictionless motion  of a particle bounded by a closed surface where it is
specularly reflected is known as {\em billiard dynamics} and dates back to Birkhoff \cite{BIR27}. 
It serves to illustrate the transition from strict regularity (integrability) to chaos (ergodicity) in 
Hamiltonian systems \cite{billiard} and bears important connections to {\em quantum chaos} as well
\cite{billiard-qc}.   We have chosen to study the 2D {\em cosine billiard} where the surface is
parametrized in polar coordinates by the relation
\be
    r(\phi) = 1 + {\epsilon}\, \cos \phi \quad .
\label{eq-cos}
\ee
The parameter $\epsilon$ is a measure of nonintegrability since for $\epsilon= 0$, the 
curve is a circle and represents the integrable limit. For all $\epsilon \neq 0$, there are finite
regions of phase space that contain chaotic trajectories. Figure (8) shows on the left the mixed
and complex structure of phase space for $\epsilon_0= 0.3$: the state variables are
 the incident angles on the surface,  $\{\alpha_n\}$ , and the polar angles of the point of impact,
$\{\phi_n\}$.  Since motion is free in between collisions with the surface, our example belongs
to the class of 2D area-preserving mappings, where attractors are absent
and replaced by stochastic bands mixed with regular regions. Within these bands, the motion is
ergodic: the blackened region is produced by {\em one} single chaotic orbit.  Embedded in this stochastic web,
one observes a number of UPOs whose physical trajectories inside the boundary are shown in the 
middle portion of Figure (8).  By pulsating the deformation parameter $\epsilon$ about its nominal
value, we have achieved control of 3 UPOs of period 4, 5 and 9. We used the MED control algorithm
with  a neighbourhood of $\delta = 10^{-2}$. A numerical OGY method gives identical performance.

We mention that the successful control of billiard dynamics may offer a solution to the
degradation of finesse in resonant optical microcavities \cite{NOC97}. It has been inferred  that
the loss of lasing activities might be associated with ray chaos (geometrical limit)  in the optical resonators
where the photons are transported (via chaotic diffusion) to regions of phase space where refractive
escape (Snell's law) becomes possible. The dielectric droplets making up the resonators
behave very much like 2D billiards and we propose that programmed variations of the asymmetry
may help reduce photon leakage. The viability of the proposal is currently being investigated.
%
\subsection*{Diamagnetic-Kepler Hamiltonian}
%
%
\begin{figure}[b]
\hfill \scalebox{0.75}{\includegraphics{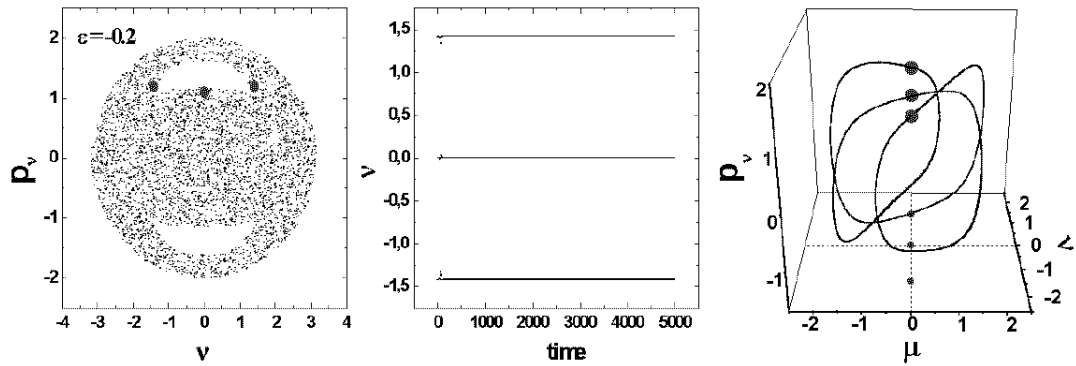}} \hfill
\vspace{10pt}
\caption{{\bf Diamagnetic Kepler Problem} for scaled energy $\epsilon_0 = -0.2$:
({\em left}) Poincar\'e section $\mu= 0, \dot{\mu} > 0$ showing one chaotic trajectory
(filled space) and an OGY controlled period 3 orbit (black dots);
({\em middle}) the stabilized $\nu$ variable on the Poincar\'e section;
({\em right}) corresponding 3D trajectory. The control settings are $\delta= 10^{-3}$ and
$\delta \epsilon_{\rm max}= 7 \times 10^{-2}$.}
\end{figure}

Our final example is a continuous, 2 degrees of freedom (4D phase space) Hamiltonian system.
It represents the motion of an electron under the combined influence of a Coulomb and a 
magnetic field. It goes under the name, {\em diamagnetic Kepler problem} (DKP), and 
occupies central stage in classical and quantum chaos research \cite{BLU97}.  We use 
scaled semi-parabolic coordinates and write the resulting scaled (pseudo-) Hamiltonian as  
(for angular momentum $L=0$)  
\be
   \hat{h}_{DK} = \frac{1}{2}(p^2_{\nu} + p^2_{\mu}) -{\epsilon}\, (\nu^2 + \mu^2)
                  + \frac{1}{8} \nu^2 \mu^2 (\nu^2 + \mu^2) \equiv 2 \quad .
\label{eq-dkh}
\ee 
The scaled energy $\epsilon$ is related to the physical energy $E$ by $\epsilon = \gamma_0^{-2/3}\ E$
where the parameter $\gamma_0 = B / B_0$ denotes the strength of the magnetic field relative
to the unit $B_0 \simeq 2.35\ 10^5\ T$.  The classical flow covers a wide range of Hamiltonian
dynamics reaching from bound, nearly integrable behaviour to completely chaotic and unbound motion
as the scaled energy is varied \cite{dkp}.

The dimension reduction (from 4D to 2D) 
and discretization is performed by observing the dynamics on the Poincar\'e section defined
by $\mu= 0, \dot{\mu}> 0$. The energy shell is then mapped to an area bounded by the
condition $p_{\nu}^2 - 2 \epsilon\ \nu^2 = 4$ which represents an ellipse in the
$(\nu, p_\nu)$ plane. The left panel of Figure (9) shows the collection of points $\{\nu_n, p_{\nu,n}\}$
obtained by numerical integration of the equations of motion for $\epsilon_0= -0.2$.
One notices, for this energy, that  phase space has few regular structures: apart from two lobes
of regularity, the rest of the ellipse is filled by the successive piercings of {\em one} chaotic trajectory.
The three dots indicate the positions of a UPO of period 3. We have succeeded in stabilizing a number
of UPOs for the system, one of them is displayed with its 3D trajectory in Figure (9). We have 
employed a complete numerical implementation of the OGY strategy.
 
In attempting to bring order to the DKP dynamics, we had to overcome a number of difficulties
not encountered in our previous examples. First, a typical trajectory spends a lot of time away
from the Poincar\'e section and because of the sensitivity of the dynamics we had to device 
an efficient variable step {\em symplectic} integrator thereby preserving the geometrical structure of the Hamiltonian.
Second, we had to obtain numerical Jacobians for all members of the UPOs since it was found to be
necessary to intervene at {\em every} crossing of the Poincar\'e section.  Third, the eigenvalues
of  area-preserving Jacobians are often complex and the stable and unstable manifolds are no longer along
the directions of their eigenvectors. A new method had to be implemented. Details of the ingenious
solutions to these problems can be found in \cite{POU99}.

We should comment that this is the first control of a realistic  Hamiltonian system. It still 
remains an open question however if manipulations of the magnetic field to induce stabilization of a 
classical unstable orbit can be extended to the control, for example,  of Rydberg wave packet dynamics. 
  
\subsection*{Properties of the Control Procedure}

The lessons learned through the previous examples and many more not reported here allow us to
draw a list of the properties and advantages of the  adopted control `philosophy' and to point to 
remaining difficulties. \\
{\ }\\
-- no model dynamics is required {\em a priori} and only {\em {local information}} is needed;\\
-- computations at each step are minimal;\\
-- {\em {gentle touch}}: the required changes in $p_0$ can be quite small ($ < 1 \%$ );\\
-- {\em {multi-purpose flexibility:}}
      different periodic orbits can be stabilized for the {\em same} system
      in the {\em same} parameter range;\\
-- control can be achieved even with imprecise measurements of
      eigenvalues and eigenvectors: the methods are robust;\\
-- the methods can also be applied to {\em {synchronization}} of several chaotic systems.

At least three complications come to mind when one considers the implementation of chaos control
strategies to the laboratory. The presence of noise, ignored so far, may induce occasional loss of control
 or hinder  it altogether. The average waiting time to fall in the $\delta$-neighbourhood
 may be very (too) long, especially for Hamiltonian systems and a targeting strategy 
 \cite{targeting} should complement the control method.   
Furthermore, the system's parameters may drift with time and this nonstationarity should be
accounted for by  updating the control informations. {\em Tracking} \cite{tracking}
is the name given to this procedure. 
\newpage
\section*{Conclusions and Future Perspectives}
\begin{flushright}
{\footnotesize 
{\em CHAOS often breeds life, \\
When  ORDER breeds habit.}\\

{\sc Henry Brooks ADAMS} }
\end{flushright}

We have presented the basic techniques for controlling chaos and we have reported on some of our
efforts to recover order from chaos. Applications of the control of chaos have been reported
in such diverse areas as aerodynamics, chemical engineering, communications, electronics,
fluid mechanics, laser physics, as well as, biology, finance (not confirmed!),  
medicine, physiology, epidemiology and the list is constantly growing.   
It is perhaps instructive at this point to quote some of the earliest {\em experimental} successes 
of the methods  to gain an idea of the breadth and diversity of the systems considered:   
in  {\em solid state devices} and {\em condensed matter}, magneto-elastic  ribbon \cite{DIT90},
spin-wave system \cite{AZE91}, electric diode \cite{HUN91}; 
in {\em fluid mechanics}, regularization  of chaotic convection \cite{SIN91};  
in {\em chemistry}, mechanisms of control of autocatalytic reactions \cite{PEN91}; 
in {\em  laser systems}, stabilization of coupled ensemble of lasers \cite{ROY92};
in {\em physiology}, cardiac arrhythmia \cite{GAR92}, (anti)-control of epileptic seizures \cite{SCH94}.

The last  decade has seen much accomplishments and challenges for the future
are numerous. The following items seem to provide a glimpse of things to come:
generalization to spatio-temporal chaos,  adaptive control for non-stationary dynamics,
effective  control in the presence of noise (dynamical and/or observational),   adaptive synchronization of chaos.

However, the greatest challenge will remain for some times the application to complex biological systems and in particular to brain dynamics\cite{GOL92}.
Despite early efforts, euphoria has been replaced by  a healthy skepticism.  Indeed, 
complex natural systems are noisy, contain a strong stochastic component and not endowed with a  behaviour called chaos (at least not in its mathematical  rigorous sense).  
Yet,  one would like to believe that ``the controlled chaos of the brain is more 
than an accidental by-product of the brain complexity''  \cite{FRE91}.  The perspective of unifying the techniques of deterministic chaos control with a statistical stochastic description as a possible therapeutic strategy against dynamical diseases  is surely something to consider.

\vspace{1.0cm}
{\bf{Acknowledgments.}} $\cal LJD$  thanks the  members of his research group for their
contributions and stimulating discussions on the subject. He is also grateful for the hospitality of
G. Deco and B. Sch\"urmann (M\"unchen) where this work took its flight and to J. Honerkamp and
J. Timmer (Freiburg) where this report was completed. 

}

%
%
\newcommand{\pre}[1] {{\it Phys.\ Rev.\  E} {\bf #1} }
\newcommand{\prl}[1] {{\it Phys.\ Rev.\  Lett.} {\bf #1} }
\newcommand{\pra}[1] {{\it Phys.\ Rev.\  A} {\bf #1} }
\newcommand{\pla}[1] {{\it Phys.\ Lett.\  A} {\bf #1} }
\newcommand{\physicad}[1] {{\it Physica\ D} {\bf #1} }
\newcommand{\jpa}[1] {{\it J.\ Phys.\ A} {\bf #1} }

\newcommand{\ibif}[1] {{\it Int.\ J.\ Bifurcation and Chaos} {\bf #1} }
\newcommand{\chaos}[1] {{\it Chaos} {\bf #1} }


\begin{references}
\footnotesize{
\bibitem{DYS88}
Dyson, F., 
{\it Infinite in All Directions}, New York: Harper and Row Publishers, 1988,  chap. 10, pp. 182-184.

\bibitem{POI1890}
Poincar\'e, H., 
{\it Acta\ Mathematica} {\bf 13}, 1 (1890). 

\bibitem{LOR63}
Lorenz, E.N.,
{\it J.\ of Atmos.\ Sci.} {\bf 20}, 130 (1963).

\bibitem{OTT90}
Ott, E., Grebogi, C.,  and  Yorke, J.A., 
  {\it Phys.\ Rev.\  Lett.} {\bf 64}, 1196 (1990).

\bibitem{DIT90}
Ditto, W.L., Rauseo, S.N.,  and Spano, M.L.,
\prl{65}, 3211 (1990).

\bibitem{REY93}
Reyl, C., Flepp, L., Baddi, R., and Brun, E., 
{\it  Phys.\ Rev.\ E} {\bf 47}, 267 (1993).

\bibitem{DRE92}
Dressler, U., and  Nitsche, G., 
{\it Phys.\ Rev.\  Lett.} {\bf 68}, 1 (1992).

\bibitem{ROL93}
 Rollins, R.W., Parmananda, P., and Sherard, P., 
{\it Phys.\ Rev.\ E} {\bf 47}, R780  (1993).

\bibitem{HUN91}
 Hunt, E.R., 
{\it  Phys.\ Rev.\ Lett.} {\bf 67}, 1953 (1991).


\bibitem{OTT94}
Ott, E., Sauer, T., and Yorke, J.A., eds.,
{\it Coping with Chaos} , New York: John Wiley  \& Sons Inc., 1994.  

\bibitem{jacobian}
Eckman, J.P., Kamphorst, S.O., Ruelle, D., and Ciliberto, S., 
\pra{34}, 4972 (1986); Sanon, M., and  Sawada, Y.,
\prl{55}, 1082 (1985).

\bibitem{embed}
Takens, F.,
{\it Lectures Notes in Math.} {\bf 898} (1981);
Sauer, T., Yorke, J.A., and Casdagli, M.,
{\it J.\ Stat.\ Phys.} {65}, 579 (1991).
 

\bibitem{targeting}
Shinbrot, T., Ott, E., Grebogi, C., and Yorke, J.A., 
\prl{65}, 3215 (1990);
Kostelich, E.J., Grebogi, C., Ott, E., and Yorke, J.A.,
\pre{47}, 305 (1993).

\bibitem{HEN76}
H\'enon, M., 
{\it Comm.\ Math.\ Phys.} {\bf 50}, 69 (1976).

\bibitem{ROS79}
R\"ossler, O.E., 
\pla{71}, 155 (1979).

\bibitem{BIR27}
Birkhoff, G.D., 
{\it Acta Math.} {\bf 50}, 359 (1927).

\bibitem{billiard}
Berry, M.V.,
{\it Eur.\ J.\ Phys.} {\bf 2}, 91 (1981); 
Robnik, M., 
\jpa{16}, 3971 (1983).

\bibitem{billiard-qc}
St\"ockmann, H.J., and Stein, J.,
\prl{64}, 2215 (1990); Stein, J., and St\"ockmann, H.J., \prl{68}, 2867 (1992).
 
\bibitem{NOC97}
N\"ockel, J.U., and Stone D., {\it Nature} {\bf 385}, 45 (1997).

\bibitem{BLU97}
Bl\"umel, R., and Reinhardt, W.P.,
{\it Chaos in Atomic Physics}, Cambridge: Cambridge Uni. Press, 1997.

\bibitem{dkp}
Delos, J.B., Knudson, S.K., and Noid, D.W.,
\pra{30}, 1209 (1984); Friedrich, H., and Wintgen, D.,
{\it Phys.\ Rep.} {\bf 183}, 37 (1989).

\bibitem{POU99}
Pourbohloul, B., {\it Control and Tracking of Chaos in Hamiltonian Systems},
Ph.D. Thesis (Universit\'e Laval, May 1999);
Pourbohloul, B., and Dub\'e, L.J.,
{\it Control and Tracking in the Diamagnetic Kepler Problem} (submitted to Phys. Rev. Lett.)
 
\bibitem{tracking}
Schwartz, I.B., Carr T.W., and Triandaff, I.,
\chaos{7}, 64 (1997).

\bibitem{AZE91}
Azevedo, A., and Rezende, S.M.,
\prl{66}, 1342 (1991).

\bibitem{SIN91}
Singer, J., Wang, Y., and  Bau, H.,
\prl{66}, 1123 (1991).

\bibitem{PEN91}
Peng, B., Petrov, V., and Showalter, K.,
{\it J.\ Phys.\ Chem.} {\bf 95}, 4957 (1991).

\bibitem{ROY92}
Roy, R., Murphy Jr., T.W., Maier T.D., Gills, Z., and Hunt, E.R., 
\prl{68}, 1259 (1992).

\bibitem{GAR92}
Garfinkel, A., Spano, M.L., Ditto, W.L., and Weiss, J.N.,  
{\it Science} {\bf 257}, 1230 (1992).

\bibitem{SCH94}
Schiff, S.J., Jerger, K., Duong, D.H., Chang, T., Spano, M.L., and Ditto, W.L., 
{\it Nature} {\bf 370}, 615 (1994).

\bibitem{GOL92}
Goldberger, A.L.,
in {\sl Applied Chaos},  J. H. Kim and J. Stringer (Eds.),  New York: John Wiley  \& Sons Inc.,
 1992, pp. 321-331. 

\bibitem{FRE91}
Freeman, W.J.,
{\it Scientific American} {\bf Feb.}, 78 (1991).
}
\end{references}
\end{document}